\documentstyle[12pt]{article}
\def\overlay#1#2{\ifmmode \setbox 0=\hbox {$#1$}\setbox 1=\hbox to\wd 0{\hss
$#2$\hss }\else \setbox 0=\hbox {#1}\setbox 1=\hbox to\wd 0{\hss #2\hss }\fi
#1\hskip -\wd 0\box 1}
\def\case#1/#2{{\textstyle{#1\over#2}}}

\def\etal{{\it et al.}}

\catcode`@=11
\newcount\@tempcntc
\def\@citex[#1]#2{\if@filesw\immediate\write\@auxout{\string\citation{#2}}\fi
  \@tempcnta\z@\@tempcntb\m@ne\def\@citea{}\@cite{\@for\@citeb:=#2\do
    {\@ifundefined
      {b@\@citeb}{\@citeo\@tempcntb\m@ne\@citea\def\@citea{,}{\bf ?}\@warning
      {Citation `\@citeb' on page \thepage \space undefined}}%
    {\setbox\z@\hbox{\global\@tempcntc0\csname b@\@citeb\endcsname\relax}%
     \ifnum\@tempcntc=\z@ \@citeo\@tempcntb\m@ne
       \@citea\def\@citea{,}\hbox{\csname b@\@citeb\endcsname}%
     \else
      \advance\@tempcntb\@ne
      \ifnum\@tempcntb=\@tempcntc
      \else\advance\@tempcntb\m@ne\@citeo
      \@tempcnta\@tempcntc\@tempcntb\@tempcntc\fi\fi}}\@citeo}{#1}}
\def\@citeo{\ifnum\@tempcnta>\@tempcntb\else\@citea\def\@citea{,}%
 \ifnum\@tempcnta=\@tempcntb\the\@tempcnta\else
  {\advance\@tempcnta\@ne\ifnum\@tempcnta=\@tempcntb \else \def\@citea{--}\fi
   \advance\@tempcnta\m@ne\the\@tempcnta\@citea\the\@tempcntb}\fi\fi}
\catcode`@=12

\font\fortssbx=cmssbx10 scaled \magstep2

\begin{document}

\vspace*{.5in}

\thispagestyle{empty}

\hbox to \hsize{
{\fortssbx University of Wisconsin - Madison}
\hfill  $\vcenter{\baselineskip14pt
                  \hbox{\bf MADPH-95-909}
                  \hbox{\bf RAL-TR-95-063}
                  \hbox{\bf hep-ph/9511459}
                  \hbox{November 1995}}$ }

\vspace{.75in}

\begin{center}
{\large\bf Top pair production with an extra gluon at the Tevatron}\\[.6in]
V.~Barger$^a$, P.G.~Mercadante$^a$ and R.J.N.~Phillips$^b$\\[.3in]
\it
$^a$Physics Department, University of Wisconsin, Madison, WI 53706, USA\\
$^b$Rutherford Appleton Laboratory, Chilton, Didcot, Oxon OX11 0QX, UK
\end{center}

\vspace{.75in}

\begin{abstract}
We calculate top pair production and decay at the Tevatron $p\bar p$
collider, with the emission of an extra gluon, and study the
corresponding $W+5$\,jet top signals including full spin correlations
in the $W \to \ell\nu$ leptonic and $W\to jj$ hadronic decays.
We study the feasibility of reconstructing $W+5$\,jet top events with
a single $b$-tag, including realistic energy resolution.
Our suggested basic procedure based on kinematic fitting
achieves about 74\% reconstruction efficiency, with 74\% of the
reconstructed events correctly classified (purity); this improves
to 82\% efficiency with 77\% purity in double-$b$-tagged events.
We suggest possible refinements, based on virtuality criteria, that
give higher purity at the cost of lower reconstruction efficiency.
\end{abstract}

\newpage

Now that top quark signals have been seen at the Tevatron, in both the
dilepton+jets and single-lepton${}+4$\,jets channels and by both the CDF
and  D0 collaborations~\cite{cdf,dzero}, it is interesting to explore other
channels where top signals may be found.  The underlying parton mechanism
for producing these signals at the Tevatron energy is dominantly
\begin{equation}
  q + \bar q \to t + \bar t \to (bW^+)(\bar bW^-),
\label{eq:qqtt}
\end{equation}
with either one or both of the $W$ bosons decaying leptonically:
$W \to \ell\nu$ ($\ell = e, \mu$). It is important to tag one or more
of the $b$-jets, by a displaced vertex or by a lepton from $b$-decay,
in order to establish the signal and discriminate against background.
For determining the top quark mass, it is preferable to study the
$(W\to\ell\nu)+4$\,jets channels, where one of the $W$-bosons decays
hadronically ($W\to jj$) and a suitably chosen three-jet combination
has invariant mass $m(bjj)\simeq m_t$, avoiding problems with
invisible neutrinos.  The principal background in the $W+4$\,jets
channel comes from the electroweak production of a single $W$-boson
plus four QCD jets~\cite{vecbos}, but with the usual acceptance cuts
and $b$-tagging this background is much smaller than the signal.

In high-$Q^2$ processes like top pair production, it is not uncommon
that additional hard QCD radiation (typically a gluon) will be
emitted, viz
\begin{equation}
  q + \bar q \to (bW^+)(\bar bW^-)g.
\label{eq:qqbWg}
\end{equation}
Here the gluon can be radiated either from the incident quarks, or
from the produced top quarks before they decay, or from subsequent
top decays into $bW$, and complete calculations have recently been
performed~\cite{oss} exploiting the MADGRAPH program~\cite{madg}.
These improve on previous calculations that omitted radiation from
top decay~\cite{ellsex}; the new results coherently combine the
effects of radiation during both production and decay processes,
together with their interference.  The radiation of gluons from
the color-disconnected process of hadronic $W$-decay ($W\to jj$) can be
ignored here, since the hadronically decaying $W$ is identified
experimentally as a dijet with invariant mass $m(jj)\simeq M_W$. In the
five-jet channel, it seems very likely that the background from
$W+5$\,QCD~jets will also be small compared to the signal, after
$b$-tagging.

\medskip
\leftline{\bf Basic reconstruction of single-tagged events}

To make use of the resulting $(W\to\ell\nu)+5$\,jet final states
in the study of top signals, some criteria must be established to
distinguish the gluon jet from the other jets.  One $b$-jet is
identified by tagging. The $W\to jj$ dijet is identified by its
invariant mass, $m(jj)\simeq M_W$ (it is extremely unlikely that
either of these is a $b$-jet).  The remaining two jets are
presumably from one gluon and one $b$-quark; we propose to identify
the gluon with the jet of lower transverse momentum $p_T$, since
the $b$-jets from $t\to bW$ have typically high $p_T$ with a Jacobian
peak at $p_T\simeq (m_t^2-M_W^2) /(2m_t) \simeq 70$~GeV in the
$t$-restframe. Finally the $W\to \ell\nu$ decay can be reconstructed
within a two-fold ambiguity using the invariant mass
constraint $m(\ell\nu)\simeq M_W$, when we attribute the missing
transverse momentum $\overlay/p_T$ in the event to the neutrino
$\left(\overlay/{\bf p}_T = {\bf p}_T(\nu)\right)$. There are now
twelve different configurations, in which a $W+5$\,jet event can be
interpreted as top pair production and decay, with a gluon emitted
either in the initial production or in the final decay process:
\begin{eqnarray}
\mbox{Class A:} && g (t\to W_{\ell\nu} b)  (t\to W_{jj}b) \,,\\
\mbox{Class B:} &&   (t\to W_{\ell\nu} bg) (t\to W_{jj}b) \,,\\
\mbox{Class C:} &&   (t\to W_{\ell\nu} b)  (t\to W_{jj}bg)\,.
\end{eqnarray}
There are four configurations in each class, corresponding to two
$W\to\ell\nu$ solutions and two different ways to pair the $b$-quarks
with $W$-bosons. Although we evaluate all the diagrams in each event,
Class~A is well represented by diagrams a,b,e,f,g shown in
Figure~\ref{fig:diagrams}, Class~B by d,f and Class~C by c,e
(in the case of $W^+\to\ell^+\nu$ leptonic decay).
We note that gluon emission from a top quark can contribute to
Class A or B or C.
The underlying idea for event reconstruction is that events are most
likely to occur in regions of phase space where one class of Feynman
diagrams has both a top propagator and an antitop propagator near the
mass shell, and are unlikely otherwise; the near-shell propagators
define the event class. Thus almost all events fall
into one of the Classes A,B,C, although a very small fraction may
defy this classification (e.g. one top may decay far off-shell).

Lepton-tagging of the $b$-jet would distinguish $b$ from $\bar b$
and hence reduce the number of competing configurations to six
(two in each class), improving the prospects for a correct
reconstruction.  Our analysis neglects this positive feature,
implicitly assuming vertex-tagging;
however, we also neglect for simplicity the negative effects
of possible mistagging (illustrated in Ref.~\cite{mistag}).

Our procedure is first to identify the gluon and other jets as
indicated above (with $W\to jj$ the best fit of untagged jet pairs),
and then to evaluate the invariant masses $m_1, m_2$ of the
two ``top" candidate clusters in each configuration, and to assign
a closeness-of-fit parameter
\begin{equation}
F = (m_1-m_t)^2 +(m_2-m_t)^2,
\label{eq:fdef}
\end{equation}
assuming that the top mass $m_t$ will have been accurately determined
from $W+4$-jet events. The configuration with lowest $F$ is designated
the best fit ; it assigns the event to Class~A, B, or~C, gives the
reconstruction of all momenta, and fixes which jet is $b$ and which
is $\bar b$. We require $F_{\rm min}<500\rm~GeV^2$ for an acceptable
fit $\left( \sqrt{F_{\rm min}/2}=16\rm~GeV\right)$, otherwise the
reconstruction is deemed to fail.

\medskip
\leftline{\bf Tests with Monte Carlo events}

We have tested this procedure with Monte Carlo $(W\to\ell\nu)+5$\,jet
events generated by the MADGRAPH program~\cite{madg}, using the observed
top mass $m_t=174$ GeV~\cite{cdf,dzero} and calculated decay width
$\Gamma_t=1.53$ GeV. To simulate detector energy resolution
we add realistic gaussian smearing:
\begin{eqnarray}
\Delta E/E &=& 0.15 \Big/ \sqrt{E/\rm GeV}\quad \mbox{(for leptons)}\,,\\
\Delta E/E &=& 0.8 \Big/ \sqrt{E/\rm GeV}\quad \mbox{(for quarks)}\,.
\end{eqnarray}
We also make the following acceptance cuts, broadly typical of
Tevatron top analyses:
\begin{equation}
\begin{array}{rcl@{\qquad}rcl}
p_T(\ell) &>& 20\rm\ GeV & |\eta(\ell)| &<& 2.5 \\
p_T(j) &>& 10\rm\ GeV & |\eta(j)| &<& 2.5 \\
\Delta R(\ell j) &>& 0.4 & \Delta R(jj) &>& 0.4 \\
\overlay/p_T &>& 25\rm\ GeV
\end{array}
\end{equation}
where $\eta = \ln\tan(\theta/2)$ is pseudorapidity, $(\Delta R)^2 = (\Delta
\eta)^2 + (\Delta \phi)^2$ measures angular separation, while $\theta$ and
$\phi$ are the usual polar and azimuthal angles with respect to the beam.
These cuts are applied at the parton level, interpreting quarks and
gluons as jets.

In the absence of smearing, we find that our procedure correctly
reconstructs about 95\% of single-$b$-tagged events that pass the
acceptance cuts; misreconstructions and failures occur in the 5\% of
events where the gluon has higher $p_T$ than the untagged $b$-jet (and
is therefore incorrectly identified). There are very few events where
the top-quark propagators are so far off-shell that they alone give
$F_{\rm min} > 500$~GeV.

For smeared Monte Carlo events, the success of our reconstruction procedure
(after acceptance cuts)  is shown in Table~\ref{reconstruct1}. The
first two columns give the percentage of events in true classes, determined
from event kinematics before smearing. Columns 3--6 show the corresponding
percentages that are reconstructed in Classes A--C or fail (because $F_{\rm
min}>500\rm~GeV^2$). Failures and misreconstructions typically arise in events
where, after smearing, the wrong pair of jets gives the best fit to $W\to jj$,
or the gluon has higher $p_T$ than the untagged $b$-jet.

\begin{table}
\caption{\label{reconstruct1}
Basic reconstruction for smeared single-$b$-tagged events after cuts.}
\smallskip
\centering
\tabcolsep2em
\begin{tabular}{crcrrrr}
\hline\hline
\multicolumn{2}{c}{\hidewidth True class and percent}\hidewidth& &
\multicolumn{4}{c}{Percentage in reconstructed classes}\\
&&& A& B& C& Fail\\
\hline
A& 59.7& $\Rightarrow$& 32.5& 4.8& 4.5& 17.9\\
B& 20.2& $\Rightarrow$&  4.1& 9.9& 2.0& 4.2\\
C& 20.1& $\Rightarrow$&  2.5&  1.6& 12.1& 3.9\\
\hline\hline
\end{tabular}
\end{table}
These results show that 55\% of events passing the cuts are correctly
reconstructed in~Class A, B or C, while 19\% are incorrectly reconstructed
(in the wrong class) and 26\% fail to reconstruct; in other words,
our procedure has 74\% reconstruction efficiency and 74\% purity
(correctness of classification), albeit with different
degrees of purity (83\%,61\%,65\%) in different reconstructed classes (A,B,C).
We surmise that a similar success rate would be achieved with real data.

It is interesting to investigate how well such reconstructed events reproduce
the correct dynamical distributions for gluons emitted before (Class~A) or
during (Classes~B,C) the top quark decay, i.e.\ whether the reconstruction
procedure introduces significant biases.
Figure~\ref{fig:p_T(g)} shows distributions versus gluon transverse momentum
$p_T(g)$ for Class~A,B,C events; solid histograms represent true unsmeared
events while dashed histograms compare the behaviour of reconstructed
smeared events (normalized to the same area). The solid/dashed
discrepancies can be qualitatively understood as follows.  Class A events
with soft gluons (hence small $p_T$) can rather easily fake
B or C after smearing, because such gluons affect invariant masses
rather little, so we lose A and gain B,C events at small $p_T(g)$.
There is a flow the other way, too, which apparently wins out at larger
$p_T(g)$.  We see that the true A,B,C $p_T(g)$-dependences are very
similar (solid histograms), but the misidentification probabilities
change with $p_T(g)$ and the dashed histograms are rather different.

Similarly, Fig.~\ref{fig:DeltaR_gb} compares true and reconstructed
distributions of the separation $\Delta R(gb)$ between the gluon jet
and its associated $b$-jet in Class~B and C events.  True events
show the expected sharp peak at small $\Delta R$ (cut off by our
acceptance criteria at $\Delta R=0.4$), due to the propagator of
the off-shell $b^*$-quark in the radiation process $b^*\to bg$.
Reconstructed histograms, however, contain 20-30\% backgrounds of
misidentified events (mostly from Class A) with different dynamical
origins that give no such peak.

Figure~\ref{fig:E(g)} compares true and reconstructed distributions
versus gluon energy $E(g)$ in the parent top rest-frame, for Class~B
and C events. The same 20--30\% backgrounds are present here too,
but apparently have much the same $E(g)$-dependence as the true
signal.

\medskip
\leftline{\bf Refined reconstruction for single-tagged events}

The results above show that many misreconstructed events do not
have the expected close correlation with the beam line (Class A, see
Fig.~\ref{fig:p_T(g)}(a)) or with the associated $b$-jet (Classes B and
C, see Fig.~\ref{fig:DeltaR_gb}).
We have therefore investigated ways to incorporate such correlations
into the reconstruction procedure; it seems that the best-motivated
way is to introduce the relevant virtuality in each configuration,
as follows.   For Class A, we consider the virtuality
$[p(q^*)]^2 = [p(q)-p(g)]^2 = -2p(q).p(g)$
of the off-shell quark $q^*$ that would be recoiling against a
gluon $g$ radiated from an initial quark or anti-quark $q$; we
choose the lowest of the two possible values corresponding to the
incident quark and anti-quark; this quantity is $\leq$ 0 and
vanishes at the $q^*$-propagator pole.
For Classes B and C, we consider the virtuality
$[p(b^*)]^2-m_b^2 = [p(b)+p(g)]^2-m_b^2 = 2p(b).p(g)$
of the off-shell $b^*$-quark that would be radiating the gluon
in these configurations; this quantity is $\geq$ 0 and vanishes
at the $b^*$-propagator pole. Clearly, a small virtuality implies a
large matrix element and hence a large likelihood that the gluon
was emitted in the corresponding configuration; this suggests
minimum-virtuality as an additional criterion in choosing the
best fit.

We note, incidentally, that minimum-virtuality alone cannot
select a unique configuration in our analysis, since each B-type
configuration has the same $b^*$-virtuality as a C-type
configuration where the $(b,W)$ pairings are interchanged;
similarly, each A-type configuration has the same $q^*$-virtuality
as another A-type with $(b,W)$ pairings reversed.  However,
in the particular case of lepton-tagging the pairings would be fixed
and this degeneracy would disappear.

Accordingly, we propose to combine kinematic fitting with a
minimum-virtuality criterion. We now accept a given configuration
as the best fit if it has both (a) the minimum $F$ with value $< 500$
and (b) the minimum absolute value of virtuality, compared to
all the competing configurations.
The results of this more refined strategy are shown in
Table~\ref{reconstruct2}.

\begin{table}
\caption{\label{reconstruct2}
Refined reconstruction for smeared single-$b$-tagged events after cuts.}
\smallskip
\centering
\tabcolsep2em
\begin{tabular}{crcrrrr}
\hline\hline
\multicolumn{2}{c}{\hidewidth True class and percent}\hidewidth& &
\multicolumn{4}{c}{Percentage in reconstructed classes}\\
&&& A& B& C& Fail\\
\hline
A& 59.7& $\Rightarrow$& 15.7& 2.5& 2.3& 39.2\\
B& 20.2& $\Rightarrow$&  0.5& 8.3& 1.1& 10.4\\
C& 20.1& $\Rightarrow$&  0.3& 1.0& 9.1&  9.7\\
\hline\hline
\end{tabular}
\end{table}

These results shows a marked improvement in purity, which is now
95\%, 70\%, 73\% in reconstructed classes A,B,C respectively
(81\% overall).
Figure~\ref{fig:DeltaR_gb2} presents the corresponding $\Delta
R(gb)$ distributions in Class~B and C events. We can see that the extra
virtuality criterion brings the reconstructed distribution much closer
to the true unsmeared case than previously (Fig.~\ref{fig:DeltaR_gb}).

However, efficiency has now dropped to about 41\% overall
and is particularly low (34\%) in events of true class A.  The reason
for the latter is that the different virtuality distributions are
affected in quite different ways by our acceptance cuts,
as we now describe.
In true class B events, the relevant virtuality $[p(b^*)]^2$
peaks at zero before cuts, but this peak is removed by the $\Delta R(bg)$
cut and the remaining events have a peak around $(18\;\rm GeV)^2$
that is not much smeared by energy resolution; in contrast, the
``wrong" virtualities (corresponding to incorrect A or C assignments)
have broader distributions peaking near $(\mbox{50--60~GeV})^2$ instead,
so the minimum-virtuality criterion
usually points to the correct B assignment.
In true class A events, the relevant virtuality $[p(q^*)]^2$ also
peaks at zero before cuts; this peak is cut out by the $p_T(g)$
and $|\eta(g)|$ cuts and the remaining distribution now
vanishes below about $(20\;\rm GeV)^2$ and has a broad shape
peaking around $(\mbox{40--50~GeV})^2$.   The ``wrong"
virtualities both have rather broad  distributions peaking near
$(60\;\rm GeV)^2$, but with wings extending down even below
$(20\;\rm GeV)^2$, so the minimum-virtuality criterion now quite
often points to an incorrect B or C assignment; increased conflict with
the minimum-F criterion gives more failures and lower efficiency.

This overlap of right and wrong virtualities happens because
the acceptance cuts act more harshly against small $[p(q^*)]^2$
than against small $[p(b^*)]^2$.    This overlap might be reduced
if there were different jet cuts,  but as things stand the
minimum-virtuality criterion is not particularly helpful in Class A
reconstructions. We therefore propose the following compromise strategy.

\medskip
\leftline{\bf Compromise strategy for single-tagged events}

Since the minimum-virtuality criterion is apparently helpful in classes
B and C, but not in class A reconstructions, a simple compromise
strategy is to apply it only in the former cases. First select the
best-fit configuration by minimizing F; if the result is Class A, accept
it; if the result is Class B or C, accept it only if it also has
minimum virtuality.  The result of this strategy is to obtain column
A from Table~\ref{reconstruct1} with columns B and C from
Table~\ref{reconstruct2}, as shown in Table~\ref{reconstruct3}.

\begin{table}
\caption{\label{reconstruct3}
Compromise reconstruction for smeared single-$b$-tagged events after cuts.}
\smallskip
\centering
\tabcolsep2em
\begin{tabular}{crcrrrr}
\hline\hline
\multicolumn{2}{c}{\hidewidth True class and percent}\hidewidth& &
\multicolumn{4}{c}{Percentage in reconstructed classes}\\
&&& A& B& C& Fail\\
\hline
A& 59.7& $\Rightarrow$& 32.5& 2.5& 2.3& 22.4\\
B& 20.2& $\Rightarrow$& 4.1 & 8.3& 1.1&  6.7\\
C& 20.1& $\Rightarrow$&  2.5& 1.0& 9.1&  7.5\\
\hline\hline
\end{tabular}
\end{table}

This gives purity 83\%, 70\%, 73\% in reconstructed classes A,B,C,
respectively .  The overall efficiency is 63\%, and is roughly the
same (62\%, 67\%, 63\%) for the three true classes A,B,C.

A caveat should now be voiced.  The measurement of final-state
$b$-quark virtualities $[p(b^*)]^2-m_b^2$ is rather straightforward,
involving just the gluon-jet and associated $b$-jet kinematics,
but initial-state virtualities $[p(q^*)]^2$ require a complete
reconstruction of the event and accumulate large uncertainties
(that have in fact been included in our calculations).  As
an alternative approach, we could choose not to rely on these
$q^*$ virtualities.  There would then be no extra constraint
on best fits of class A, just as in the compromise strategy
above.  In class B and C configurations, we could exploit the
more accessible $b^*$ virtualities by simply requiring them
to be small, say less than $(50\;\rm GeV)^2$.  This {\it ad hoc}
prescription gives results very similar to Table~\ref{reconstruct3}.

\medskip
\leftline{\bf Reconstruction pattern for double-b-tagged events}

Finally it is interesting to ask how well our reconstruction strategies
would work in $W+5$-jet events where both $b$-jets have been
correctly tagged.
There is still some uncertainty here, after energy smearing, because the
best $W\to jj$ candidates may not be the correct pair of jets,
the $W\to\ell\nu$ reconstruction is still ambiguous, and we
still do not know which is the $b$-jet and which is the $\bar b$-jet
(neglecting possible lepton-tagging information as before).
Table~\ref{reconstruct4} shows the results of applying our basic
reconstruction strategy (the same as for Table~\ref{reconstruct1})
to such events.

\begin{table}
\caption{\label{reconstruct4}
Basic reconstruction for smeared double-$b$-tagged events after cuts.}
\smallskip
\centering
\tabcolsep2em
\begin{tabular}{crcrrrr}
\hline\hline
\multicolumn{2}{c}{\hidewidth True class and percent}\hidewidth& &
\multicolumn{4}{c}{Percentage in reconstructed classes}\\
&&& A& B& C& Fail\\
\hline
A& 59.7& $\Rightarrow$& 38.7&  4.6&  4.5& 11.9\\
B& 20.2& $\Rightarrow$&  4.0& 11.1&  1.8&  3.3\\
C& 20.1& $\Rightarrow$&  2.4&  1.4& 13.6&  2.7\\
\hline\hline
\end{tabular}
\end{table}

\medskip
\leftline{\bf Summary}
Our results may be summarized as follows.\\
(1) We have proposed strategies to reconstruct $(W\to\ell\nu)+5$-jet
events, that originate from $t\bar t$ pair production with the
emission of an extra gluon.  To this end, we have introduced
classes A,B,C of final states, characterized by gluon
emission in the process of (A) $t\bar tg\;$ production or
(B) $t\to W_{\ell\nu}bg\;$ decay or (C) $t\to W_{jj}bg\;$ decay.
Although in principle these classes of event must overlap, in
practice most events are expected to fall preferentially into one of
these classes (with its implied kinematical constraints); however,
a small fraction are expected to defy this classification and
thereby to escape from reconstruction.\\
(2) We have tested these strategies using Monte Carlo events
generated through the MADGRAPH program~\cite{madg}.
Since we use the full matrix elements in our calculations, the
gluon can be emitted from anywhere and the amplitudes receive
contributions from all three regions A,B,C. The distinction
between these different regions is only made in the
reconstruction procedure, where it is assumed that
a single region dominates for any given event.\\
(3) For single-b-tagged events, distinguishing the gluon from the
second $b$-jet by its generally lower $p_T$,
there are twelve competing reconstructions
(four in each class).  Our basic strategy is to select the configuration
that gives the best kinematical fit to the two reconstructed
top-quark invariant masses, i.e. minimum $F$ subject to $F_{min}<500$.
The results, shown in Table~\ref{reconstruct1}, give 74\% purity
with 74\% efficiency overall.
Figures~\ref{fig:p_T(g)},\ref{fig:DeltaR_gb},\ref{fig:E(g)}
compare some distributions of true and reconstructed events.\\
(4) A more refined strategy, where not only the top-mass discrepancies
but also the implied virtuality of the radiating beam-quark (class A)
or $b$-quark (classes B,C) are minimized simultaneously,
gives 81\% purity with 41\% efficiency, as shown in Table~\ref{reconstruct2}
and Fig.~\ref{fig:DeltaR_gb2}.  This strategy is rather wasteful in Class A
events, however, because our acceptance cuts suppress small
$q^*$-virtualities much more severely than small $b^*$-virtualities.\\
(5) A compromise strategy, where minimum-virtuality is required only
for a best fit of Class B or C, gives better efficiency 63\%
(approximately the same for all classes) while still preserving
reasonable purity 79\%; see Table~\ref{reconstruct3} and
Fig.~\ref{fig:DeltaR_gb2}.
If initial-state virtualities prove unworkable, this strategy can be
adapted to use final-state virtualities only, with similar results.\\
(6) Somewhat better results are obtained in events where both $b$-quarks
are tagged, and hence the gluon is more cleanly distinguished,
as shown in Table~\ref{reconstruct4}. Here our basic
reconstruction strategy gives 77\% purity with 82\% efficiency
(compare Table~\ref{reconstruct1} for the same strategy with
single-tagging).\\
(7) These double-tagged results are nonetheless quite far from
perfect, showing that gluon identification is only one part of the
problem. Detector resolution is responsible for essentially all the
misreconstructions in the double-tagged case, and must be a major
factor in the single-tagged case too.  Better resolution would allow
better reconstruction.

\newpage

\begin{flushleft}{\bf Acknowledgments}\end{flushleft}
We thank Erwin Mirkes and Tim Stelzer for discussions.  RJNP thanks the
University of Wisconsin for hospitality during part of this work.
VB thanks the Institute
for Theoretical Physics at the University of California, Santa Barbara for
hospitality during part of this work.
This research was supported in part by the U.S.~Department of Energy
under Grants No.~DE-FG02-95ER40896 and No.~DE-FG02-84ER40173,
in part by the National Science Foundation under Grant No.~PHY94-07194,
in part by the University of Wisconsin Research Committee with funds
granted by the Wisconsin Alumni Research Foundation and in part by Conselho
Nacional de Desenvolvimento Cient\'{\i}fico e Tecnol\'ogico (CNPq).


\newpage
\leftline{\bf Figure Captions}

\begin{enumerate}
\item{Typical Feynman diagrams for $t\bar t$ production and decay
 with an extra gluon, at the Tevatron.}
\label{fig:diagrams}

\item{Distributions versus gluon transverse momentum for true
unsmeared events (solid histograms) and reconstructed smeared
events with a single $b$-tag (dashed histograms), using our basic
reconstruction procedure;(a) Class A, (b) Class B, and (c) Class C.}
\label{fig:p_T(g)}

\item{Dependence on the separation $\Delta R(gb)$ for (a) Class B
and (b) Class C events. Solid (dashed) histograms denote true
(basic-reconstructed) events}.
\label{fig:DeltaR_gb}

\item{Dependence on the gluon energy $E(g)$ in the parent top
rest-frame for (a) Class B and (b) Class C events. Solid (dashed)
histograms denote true (basic-reconstructed) events}.
\label{fig:E(g)}

\item{Dependence on the separation $\Delta R(gb)$ for (a) Class B
and (b) Class C events. Solid (dashed) histograms denote true
(refined-reconstructed) events.  The compromise reconstruction
method also gives the dashed curves}.
\label{fig:DeltaR_gb2}
\end{enumerate}

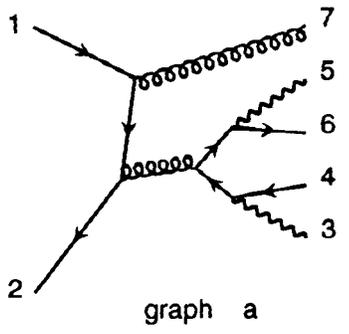
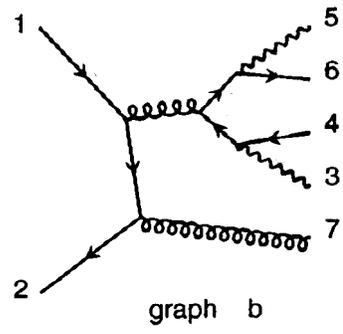
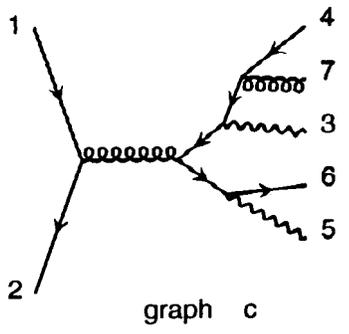
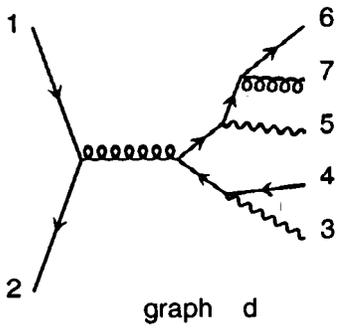
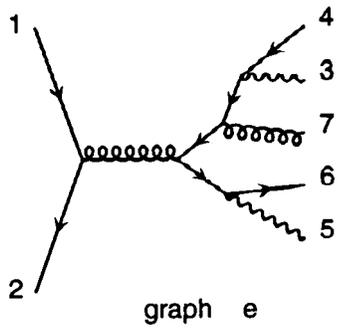
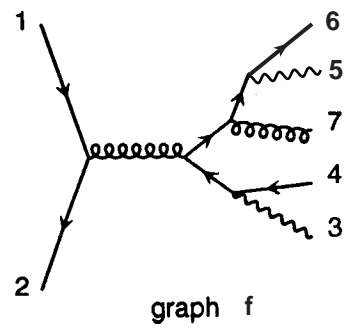
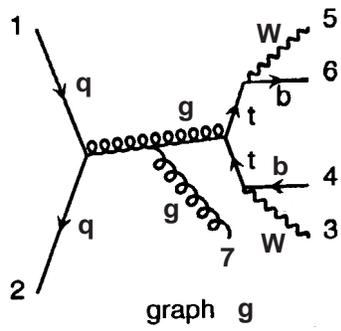
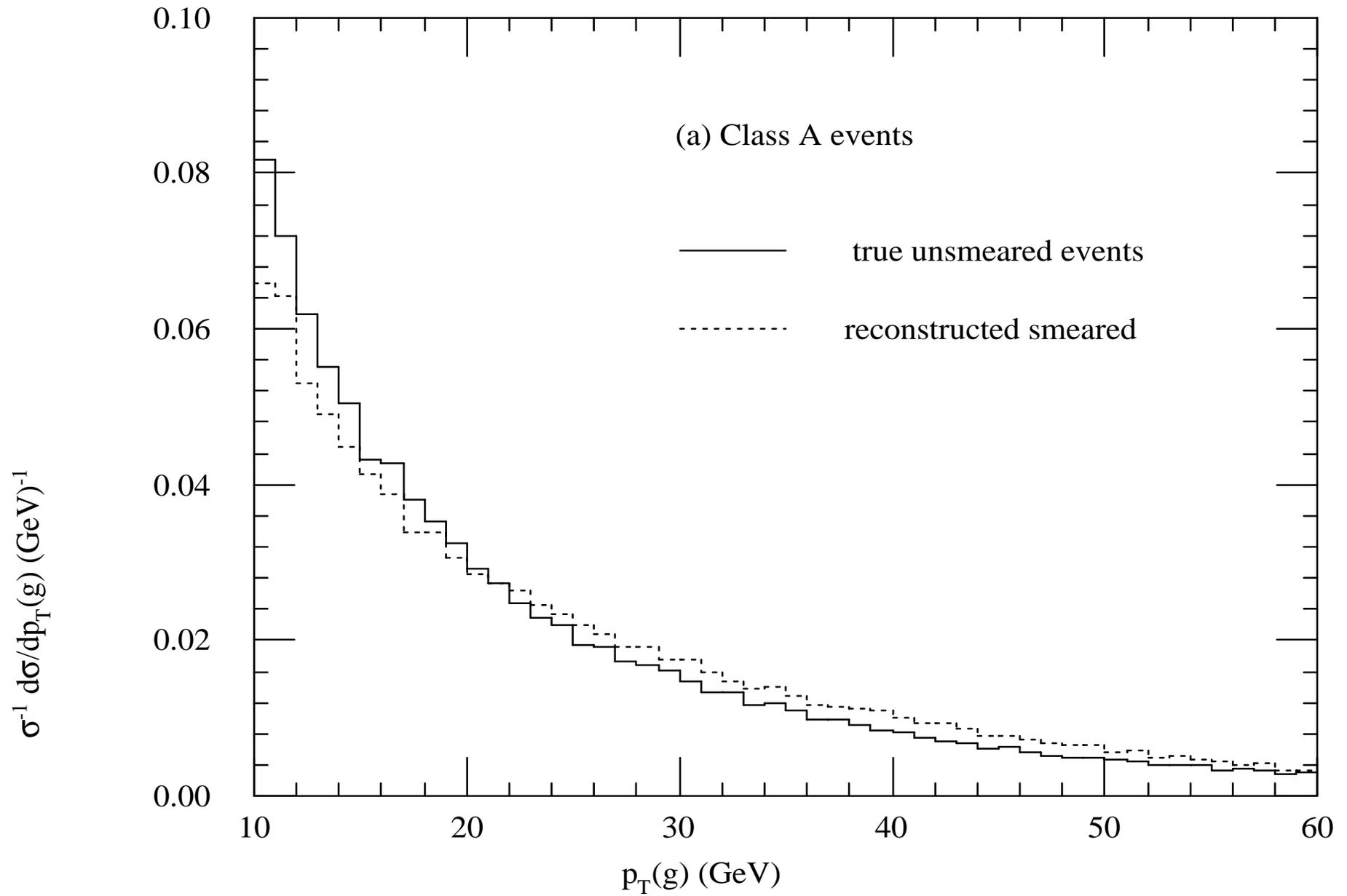
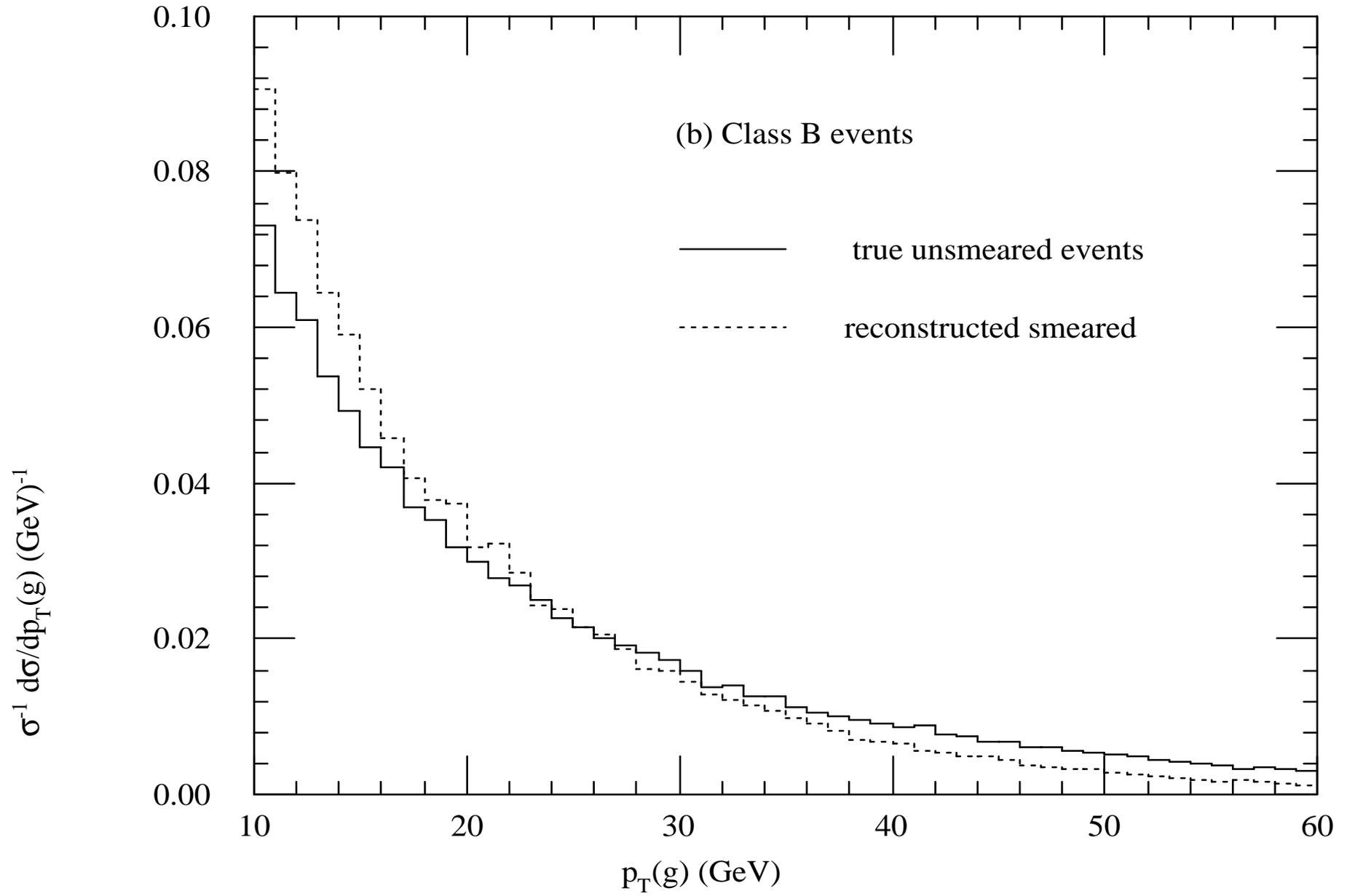
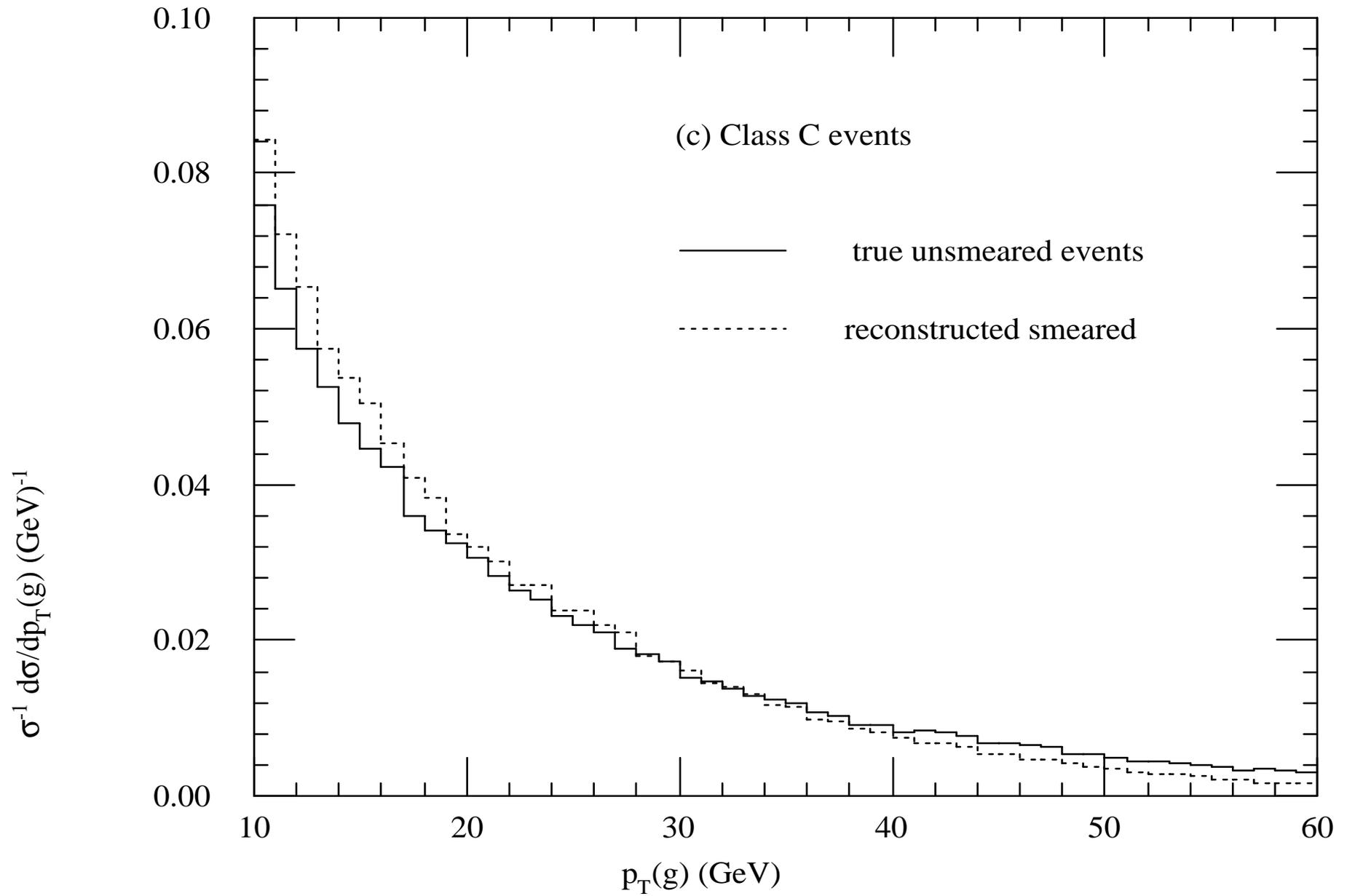
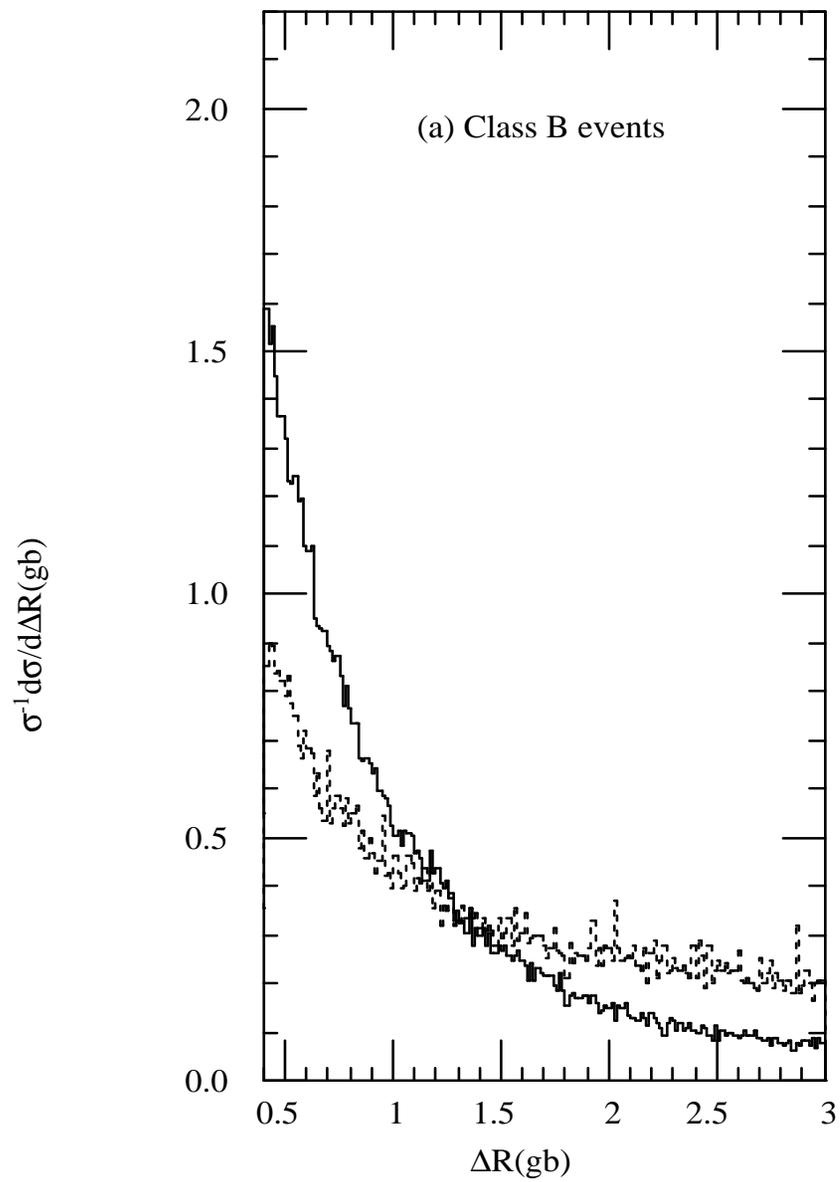
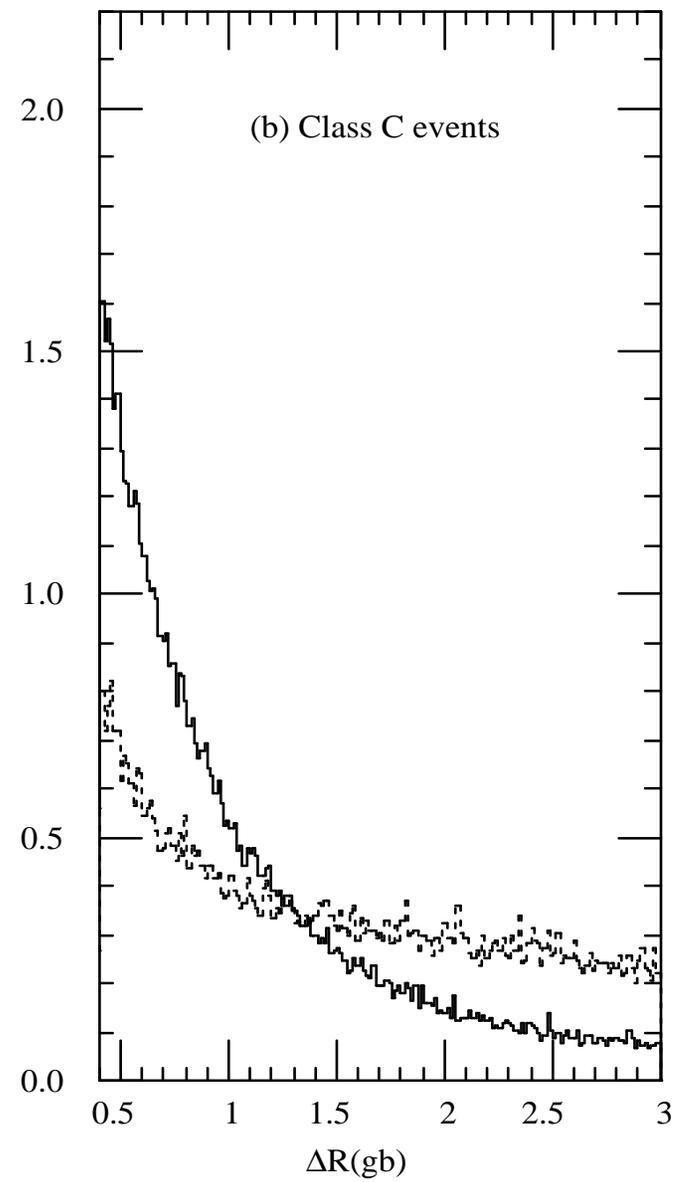
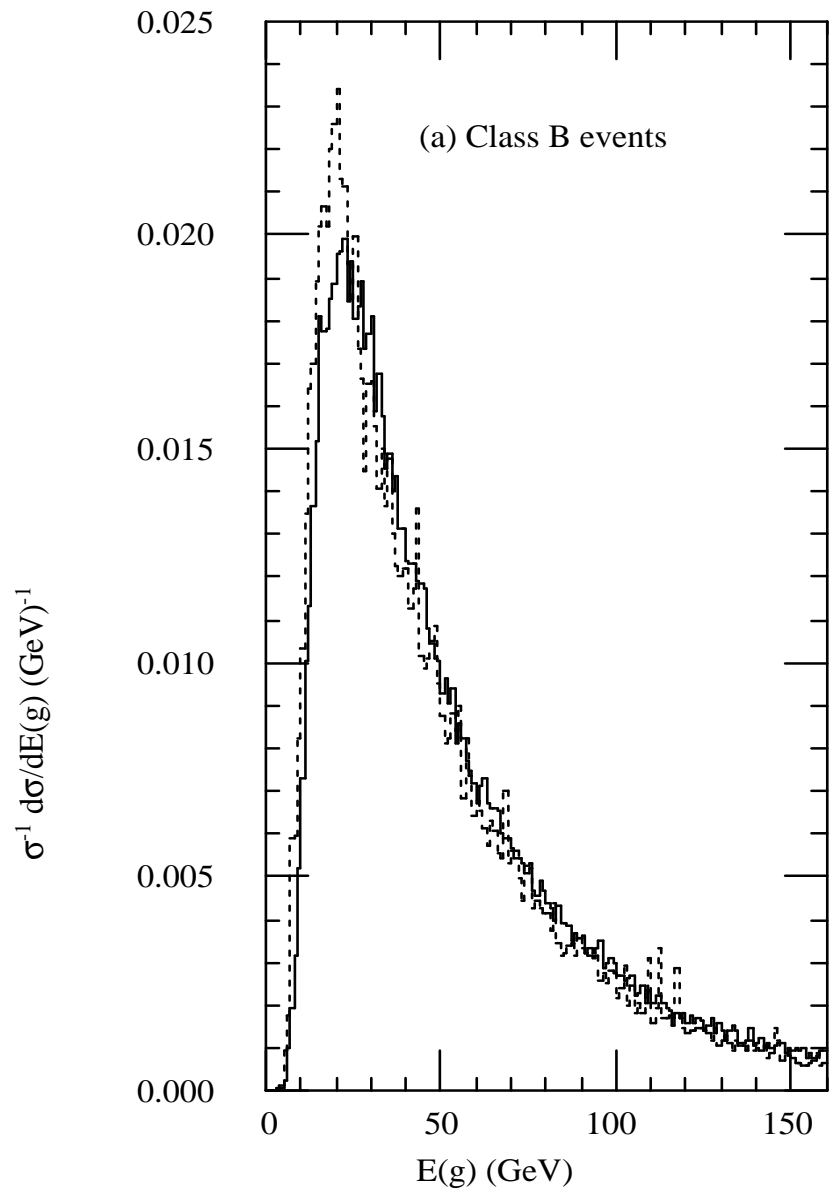
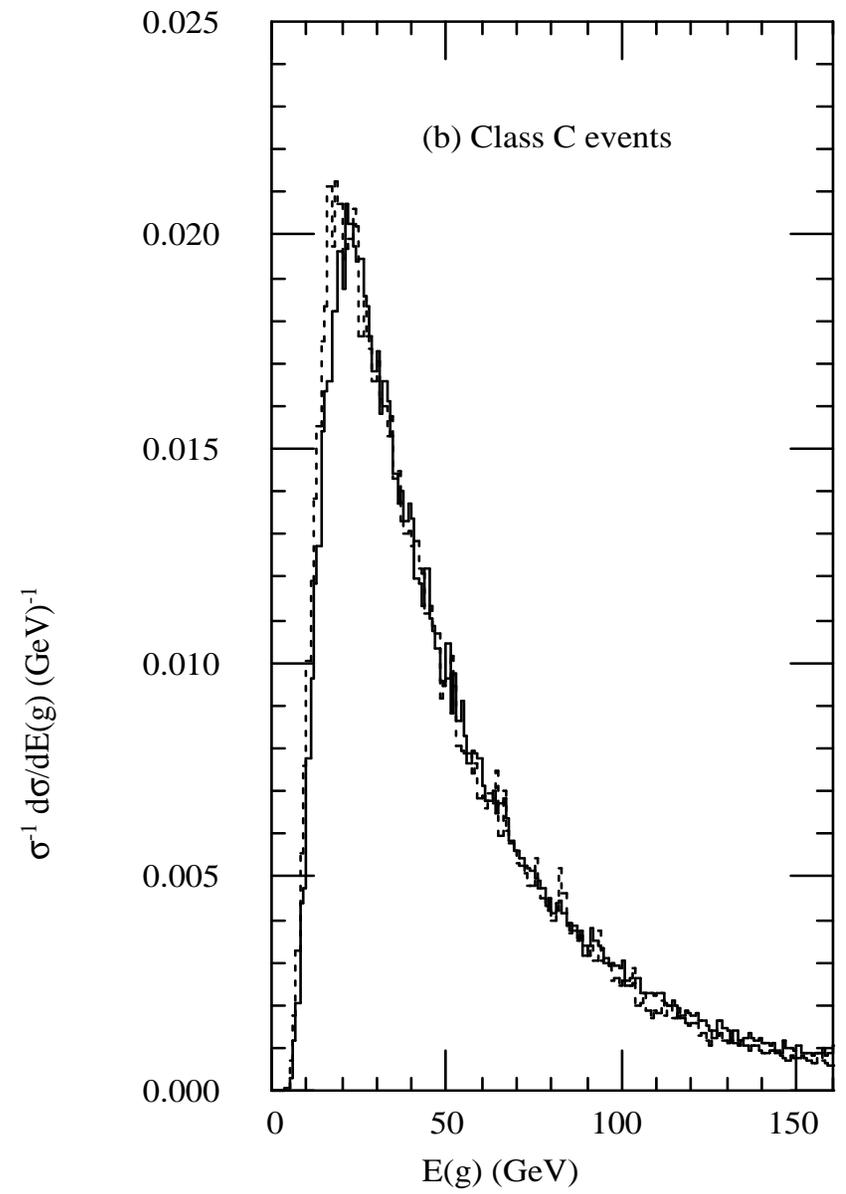
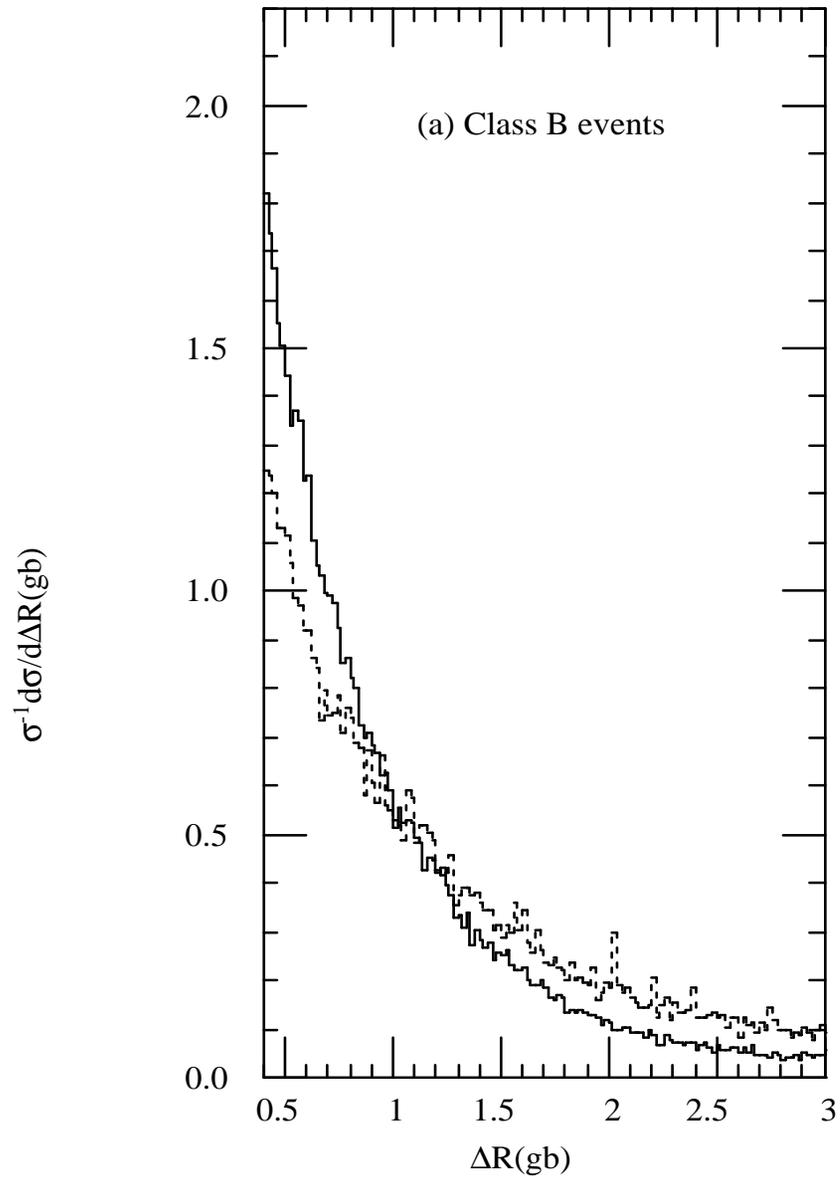
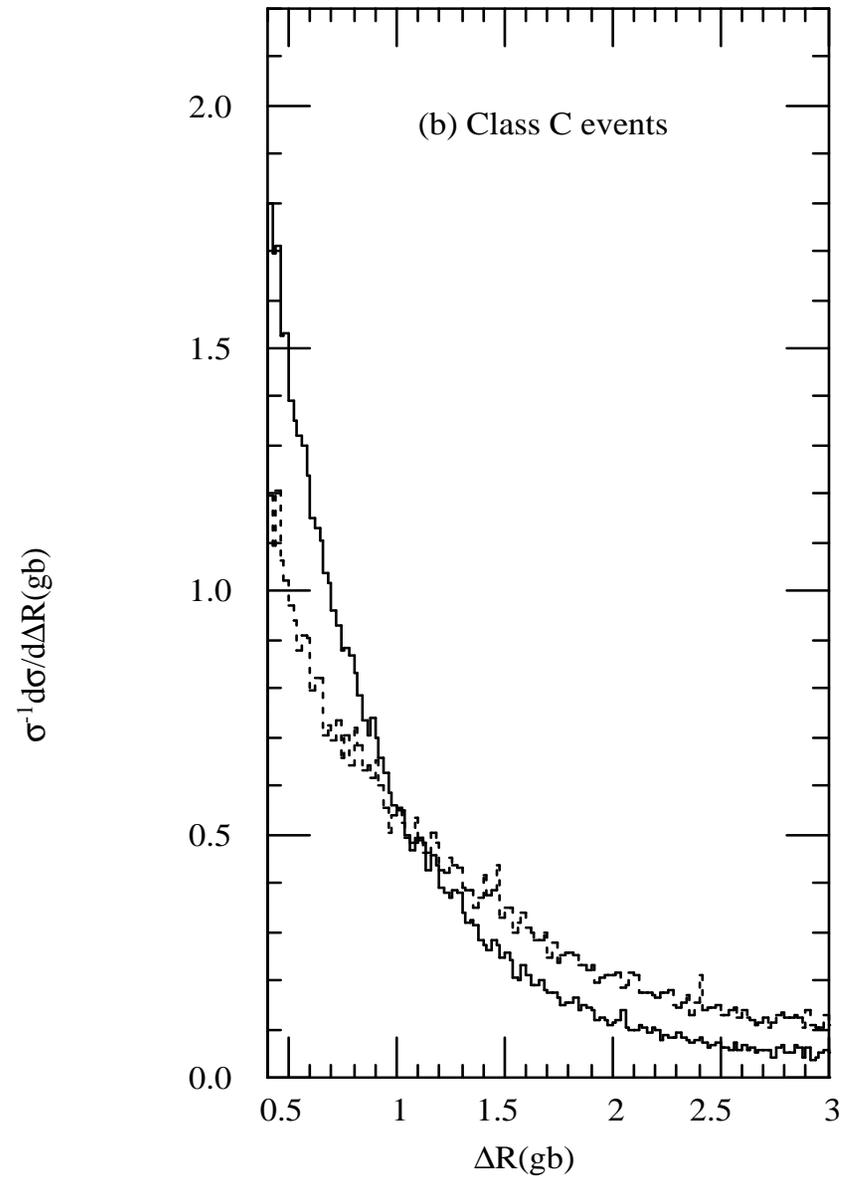
\end{document}